# Dispersion relations of the powers of complex reflection coefficient in testing the validity of THz spectra

K.-E. Peponen[a], E. Gornov, and Yu. Svirko

Department of Physics, University of Joensuu, P.O. Box 111, FI –80101 Joensuu, Finland

Y. Ino and M. Kuwata-Gonokami

Department of Applied Physics, University of Tokyo, Bunkyo-ku, Tokyo 113-8656, Japan

V. Lucarini

Department of Mathematics and Computer Science, University of Camerino, 62032 Camerino (MC), Italy

Kramers-Kronig type dispersion relations for integer powers of complex reflection coefficient are introduced for testing the consistency of terahertz reflection spectra. By using numerical simulations we show that such dispersion relations can be applied for distillation from data with some experimental artifacts without data extrapolations beyond the measured spectral range. These dispersion relations, due to causality, provide a powerful and yet uncommon tool to examine the consistency of the spectroscopic data obtained in reflection spectroscopy at terahertz range. In particular we show that real and imaginary parts of the complex reflection coefficient obtained from raw data with systematic phase error caused by sample misplacement, not necessarily obey dispersion relations, while the ones corrected with maximum entropy method obey these relations.

78.20.Ci; 78.47.+p

---

[a] Corresponding author, e-mail:kai.peiponen@joensuu.fi

## I.   Introduction

In optical spectroscopy, Kramers-Kronig (K-K) dispersion relations [1] have for a long time been utilized in data inversion. For example, they are conventionally employed to recover the wavelength dependence of the refractive index from the measured absorption spectrum of the medium. Similarly the K-K technique allow one to obtain complex refractive index from reflection spectra using a phase retrieval procedure [1]. Actually, K-K relations hold for a wide class of optical functions, e.g. one can apply them to the powers of optical constants. In this paper we focus on the powers of the complex reflection coefficient. In 1981, Smith and Manogue [2] introduced dispersion relations and relevant sum rules for the powers of complex reflection coefficient. However, these S&M dispersion relations have not attracted so much attention because they simultaneously involve both amplitude and phase of the complex reflection coefficient and hence, can not be employed to obtain complex refractive index from the conventional reflectance data. S&M dispersion relations may be useful in ellipsometry when the wavelength dependence of both phase and magnitude of the complex reflection coefficient can be measured. Similarly, S&M dispersion relations have importance in the time domain THz spectroscopy that provides us both the amplitude and the phase of the signal field. In the actual experiments, measured data have various experimental systematic errors, which may lead to experimental artifacts on the complex response functions. For example, one call recall the incorrect positioning of the sample in reflection geometry [3] or the frequency dependent loss caused by the diffraction from the edge of the sample, which is inevitable in THz region where the sample size and the wavelength are comparable. Recently, we have successfully developed an algorithm to distillate data removing phase errors using maximum entropy method (MEM) [3]. Nevertheless, MEM is a purely mathematical tool that has no physical basis. Therefore it is important to verify the quality of experimental data relying on physical principles such as causality to confirm the consistency of the complex response functions obtained from the THz experimental band-limited data.

It is necessary to notice, however, that even in optical spectroscopy, the K-K analysis may fail when one needs to retrieve the wavelength dependence of the refractive index outside the measured spectral range [4]. This problem still persists in commercial software packages supplied with spectrophotometers [5]. Very recently a numerical method to improve the accuracy of K-K analysis has been suggested [6]. Alternatively, the problem of data extrapolations beyond the measured range can be relaxed to a great extent by replacing the conventional K-K relations by multiply subtractive K-K (MSKK) relations [1], which make the analysis and inversion of the spectra more reliable both in linear and in nonlinear optical spectroscopy.

In this paper we show that the approach based on the S&M and MSKK dispersion relations has advantages for the analysis of experimental THz reflection spectra in comparison with conventional K-K technique. Specifically, when the reflection measurements are performed in a finite spectral range, the data inversion is successful without data extrapolations beyond the measured range. This is due to the fast convergence of the relevant dispersion integrals. Moreover, the S&M dispersion relations do not require a priori information on the reflection spectrum in so-called anchor points, which is an essential and sometimes also complicating feature of the MSSK analysis. In this paper we examine the applicability of the S&M and singly subtractive K-K relations (SSKK) for a case of a theoretical model, namely the band limited Lorentzian reflectivity for insulators, and also experimental THz data for a semiconductor. In particular, we find that both methods are powerful tools to detect phase error due to the sample misplacement in the THz reflection measurements.

## II. Dispersion relations for the powers of the complex reflection coefficient.

The K-K relations are based on the causality principle [1] and assumption that in the upper half of complex frequency plane (poles are located in the lower half plane) the complex reflection coefficient $r(\omega) = |r(\omega)|\exp\{i\varphi(\omega)\}$ (or complex refractive index $N(\omega)$) is a holomorphic [1] function of frequency, having a fast enough fall–off at high frequencies [1]. The assumption of the function to be holomorphic in the upper half plane is valid if the incident electric filed oscillate proportional $\exp\{-i\omega t\}$. Since the function $r^n(\omega)$ is also holomorphic for any positive integer $n = 1,2,...$, this function satisfies the following pair of S&M dispersion relations [2]

$$r^n(\omega')\cos[n\varphi(\omega')] = \frac{2}{\pi} P\int_0^\infty \frac{\omega r^n(\omega)\sin[n\varphi(\omega)]}{\omega^2 - \omega'^2} d\omega$$

$$r^n(\omega')\sin[n\varphi(\omega')] = -\frac{2\omega'}{\pi} P\int_0^\infty \frac{r^n(\omega)\cos[n\varphi(\omega)]}{\omega^2 - \omega'^2} d\omega \quad (1)$$

where P stands for the Cauchy principal value. One can observe that in relations (1), amplitude and phase are mixed up. However, the problem of data extrapolation beyond the measured spectral range is not as severe as that for the conventional K-K relations, which are presented in terms of $\log r(\omega)$ and $\varphi(\omega)$ [1]. Fast fall-off of the power of the reflection coefficient at high frequency is a crucial property for strong convergence of the dispersion relations given in equation (1).

### III. Multiply subtractive Kramers-Kronig relations for the powers of the complex reflection coefficient.

A singly subtractive K-K (SSKK) relation for the phase retrieval from the logarithm of reflectance was first employed by Ahrenkiel [7]. MSKK relations have been obtained by Nussenzveig [8] and Palmer *et al* [9] using Ahrenkiel's ideas, i. e. subtracting conventional K-K relations from each other. In particular, in [9], MSKK relation for the phase angle of the complex reflection coefficient was derived with a cumbersome calculation based on the Hermite reminder theorem and complex analysis. Below we reproduce this result using the theorem of residues. To the best of our knowledge such a derivation of MSKK relations has not been presented in the literature of this field until now.

Let us consider a holomorphic function $f(z) = u + iv$ in the upper complex half plane $\text{Im}(z) \geq 0$. We assume that $f$ tends to zero as the modulus of the complex variable tends to infinity. In dispersion theory, $z$ and $f$ may correspond e.g. to complex frequency and complex reflection coefficient or nonlinear susceptibility, respectively. Consider the case of $Q$ different anchor points $x_Q$ where the function is *a priori* known. Since $f$ is a holomorphic function in the upper half plane, Cauchy's integral theorem [4] yields

$$\oint_C \frac{f(z)}{(z-x')(z-x_1)\cdots(z-x_Q)} dz = 0, \tag{2}$$

where the closed integration contour $C$ is shown in Fig. 1. The integration can be performed piecewise along the real axis as a Cauchy principal value, along detours around simple poles $x'$, $x_1$, …, $x_Q$, and the contour is closed with a large semicircle in the upper half plane. The integral along the large semi circle vanishes according to the Jordan's lemma [4]. The principal value integral is left alone since the integrals around the small detours (their radius tends to zero in taking the principal value) provide us the residues of the integrand as follows:

$$P\int_{-\infty}^{\infty} \frac{f(x)}{(x-x')(x-x_1)\ldots(x-x_Q)} dx = i\pi\left[\frac{f(x')}{(x'-x_1)\cdots(x'-x_Q)} + \cdots + \frac{f(x_n)}{(x_Q-x')\cdots(x_Q-x_{Q-1})}\right] \tag{3}$$

Equation (3) is the origin of the MSKK relations. If the real and imaginary parts of the function $f$ have a definite parity, then one can reduce the integration interval down to $0 \leq x \leq \infty$ and arrive, for example, at the MSKK dispersion relation in terms of the phase angle of the complex reflection coefficient, given by Palmer *et al* [9].

Before proceeding it is worth to mention papers by Nash et. al [10] and Lee & Sindoni [11]. They have remarked that in the phase retrieval a constant phase correction has to be added to the phase, which is calculated from a K-K relation for the logarithm of the reflectivity. In practice, the SSKK relation for the phase angle is much more efficient and reliable tool for the data analysis than the conventional K-K relation. Since the SSKK relations are based on the subtraction one K-K relation from another, this constant phase shift is apparently cleared out. However, in order to arrive at doubly subtractive K-K relations (DSKK) one needs to subtract three K-K relations, i.e. the problem with constant phase factor reappears in the context of treating the logarithm of the reflectivity. Similarly any MSKK relation of an odd order (e. g. SSKK, TSKK etc.) allows one to obtain phase of the complex reflection coefficient that is free from the unknown phase shift. These relations essentially employ the fact that the complex reflection coefficient is *a priori* known at anchor points. This is the major difference between conventional K-K and MSKK analysis.

In order to arrive at the SSKK relations we assume that function $f = u + iv$ is *a priori* known at a single anchor point $x_1$. Separating the real and imaginary parts in Eq. (3) and assuming that the $u$ and $v$ are even and odd functions of $x = \mathrm{Re}(z)$, we find that

$$u(x') - u(x_1) = \frac{2(x'^2 - x_1^2)}{\pi} P\int_0^\infty \frac{xv(x)dx}{(x^2 - x'^2)(x^2 - x_1^2)}$$
$$v(x_1) - v(x') = \frac{2(x' - x_1)}{\pi} P\int_0^\infty \frac{(x^2 + x'x_1)u(x)dx}{(x^2 - x'^2)(x^2 - x_1^2)}$$
(4)

By substituting $u = \log|r|$, $v = \arg(r)$ and $x_1 = 0$ into (4) one can readily gain the result of Nash *et al* for the correction of the phase angle (Eq. (2) in [10]). Similarly, substitution of $u = \log|r|$, $v = \arg(r)$, $n = 1$ and $x_1 = 0$ in Eqs. (2, 3) returns the result of Lee & Sindoni ( Eq. (1) in [11]).

MSKK relations for the powers of the complex reflection coefficient $r^n(\omega)$ can be obtained from Eqs. (2, 3) by accounting the parity of the real and imaginary parts of $r(\omega)$. If the amplitude and phase of $r(\omega)$ are *a priori* known at several positive anchor point frequencies $\omega_j$, $j=1,...Q$ one can derive the following expressions:

$$|r(\omega')|^n \cos[n\varphi(\omega')]$$
$$= \frac{(\omega'^2 - \omega_2^2)(\omega'^2 - \omega_3^2)\cdots(\omega'^2 - \omega_Q^2)}{(\omega_1^2 - \omega_2^2)(\omega_1^2 - \omega_3^2)\cdots(\omega_1^2 - \omega_Q^2)}|r(\omega_1)|^n \cos[n\varphi(\omega_1)] + ...$$
$$+ \frac{(\omega'^2 - \omega_1^2)\cdots(\omega'^2 - \omega_{j-1}^2)(\omega'^2 - \omega_{j+1}^2)\cdots(\omega'^2 - \omega_Q^2)}{(\omega_j^2 - \omega_1^2)\cdots(\omega_j^2 - \omega_{j-1}^2)(\omega_j^2 - \omega_{j+1}^2)\cdots(\omega_j^2 - \omega_Q^2)}|r(\omega_j)|^n \cos[n\varphi(\omega_j)]$$
$$... + \left[\frac{(\omega'^2 - \omega_1^2)(\omega'^2 - \omega_2^2)\cdots(\omega'^2 - \omega_{Q-1}^2)}{(\omega_Q^2 - \omega_1^2)(\omega_Q^2 - \omega_2^2)\cdots(\omega_Q^2 - \omega_{Q-1}^2)}\right]|r(\omega_Q)|^n \cos[n\varphi(\omega_Q)]$$
$$+ \frac{2}{\pi}\left[(\omega'^2 - \omega_1^2)(\omega'^2 - \omega_2^2)\cdots(\omega'^2 - \omega_Q^2)\right] P\int_0^\infty \frac{\omega |r(\omega)|^n \sin[n\varphi(\omega)] d\omega}{(\omega^2 - \omega'^2)(\omega^2 - \omega_1^2)\cdots(\omega^2 - \omega_Q^2)}$$

$$\frac{|r(\omega')|^n \sin[n\varphi(\omega')]}{\omega'}$$
$$= \frac{(\omega'^2 - \omega_2^2)(\omega'^2 - \omega_3^2)\cdots(\omega'^2 - \omega_Q^2)}{(\omega_1^2 - \omega_2^2)(\omega_1^2 - \omega_3^2)\cdots(\omega_1^2 - \omega_Q^2)}\frac{|r(\omega_1)|^n \sin[n\varphi(\omega\ )]}{\omega_1} + ...$$
$$+ \frac{(\omega'^2 - \omega_1^2)\cdots(\omega'^2 - \omega_{j-1}^2)(\omega'^2 - \omega_{j+1}^2)\cdots(\omega'^2 - \omega_Q^2)}{(\omega_j^2 - \omega_1^2)\cdots(\omega_j^2 - \omega_{j-1}^2)(\omega_j^2 - \omega_{j+1}^2)\cdots(\omega_j^2 - \omega_Q^2)}\frac{|r(\omega_j)|^n \sin[n\varphi(\omega_j)]}{\omega_j}$$
$$... + \left[\frac{(\omega'^2 - \omega_1^2)(\omega'^2 - \omega_2^2)\cdots(\omega'^2 - \omega_{Q-1}^2)}{(\omega_Q^2 - \omega_1^2)(\omega_Q^2 - \omega_2^2)\cdots(\omega_Q^2 - \omega_{Q-1}^2)}\right]\frac{|r(\omega_Q)|^n \sin[n\varphi(\omega_Q)]}{\omega_Q}$$
$$- \frac{2}{\pi}\left[(\omega'^2 - \omega_1^2)(\omega'^2 - \omega_2^2)\cdots(\omega'^2 - \omega_Q^2)\right] P\int_0^\infty \frac{|r(\omega)|^n \cos[n\varphi(\omega)] d\omega}{(\omega^2 - \omega'^2)(\omega^2 - \omega_1^2)\cdots(\omega^2 - \omega_Q^2)}$$

(5)

As far as the authors know this is the first time that MSKK relations are presented for the powers of the complex reflection coefficient, however, similar dispersion relations for the moments of nonlinear susceptibility were given by Lucarini *et al* [12]. One may expect that since the denominator of the integrand in (5) depends on frequency as $\omega^{-2Q-2}$ while that in (1) as $\omega^{-2}$, the MSKK relations should benefit from faster conversion in comparison with the S&M dispersion relations (1). However, this may not be the case in data analysis of the powers of the reflection coefficient as it will be shown later. According to Palmer *et al* the anchor points should be chosen using the zeros of Chebychev polynomial of first kind. The zeros appear at [9]

$$\omega_{zero} = \left\{ \frac{(\omega_u^2 + \omega_l^2)}{2} + \frac{(\omega_u^2 - \omega_l^2)}{2} \cos\left[\frac{(2q+1)\pi}{2Q}\right] \right\}^{1/2}, \tag{6}$$

where the spectra is measured at the range $\omega_l \leq \omega \leq \omega_u$. Even at $Q = 1$, i.e. for the SSKK relations, Eqs. (5) appear to be effective for the analysis of band limited spectral data. From Eqs. (5) we get SSKK relations that can be given by expressions

$$|r(\omega')|^n \cos[n\varphi(\omega')] = |r(\omega_1)|^n \cos[n\varphi(\omega_1)] + \frac{2(\omega'^2 - \omega_1^2)}{\pi} P\int_0^\infty \frac{\omega |r(\omega)|^n \sin[n\varphi(\omega)]}{(\omega^2 - \omega'^2)(\omega^2 - \omega_1^2)} d\omega$$

$$\frac{|r(\omega')|^n \sin[n\varphi(\omega')]}{\omega'} = \frac{|r(\omega_1)|^n \sin[n\varphi(\omega_1)]}{\omega_1} - \frac{2(\omega'^2 - \omega_1^2)}{\pi} P\int_0^\infty \frac{|r(\omega)|^n \cos[n\varphi(\omega)]}{(\omega^2 - \omega'^2)(\omega^2 - \omega_1^2)} d\omega \tag{7}$$

For the sake of simplicity we have chosen the same anchor point in both dispersion relations (7).

The SSKK dispersion relations of Eq. (7) allow one to check consistency of the experimental reflection data in a similar way as S&M dispersion relations. Specifically, by using the measured amplitude and phase of the complex reflection coefficient one can calculate the corresponding real and imaginary parts from Eqs. (1) & (7) at frequency $\omega'$ that belongs to the frequency range in question. A large discrepancy between the calculated and measured complex reflection coefficient indicates a presence of an experimental error. This error may originate, for example, from the sample misplacement (spatial precision of few microns) in the time resolved THz reflection spectroscopy [3].

## IV. Numerical simulations

In order to demonstrate the performance of the dispersion relations (1) & (7) we first carry out numerical simulations using a single resonance Lorentz model [1] for the complex relative permittivity $\varepsilon(\omega)$ of a dielectric:

$$\varepsilon(\omega) = 1 + \frac{A}{\omega_0^2 - \omega^2 - i\Gamma\omega}, \tag{8}$$

where $A$ is a constant, $\omega_0$ and $\Gamma$ are resonance frequency and damping factor, respectively. At normal incidence, the complex reflection coefficient is given by

$$r(\omega) = \frac{\sqrt{\varepsilon(\omega)} - 1}{\sqrt{\varepsilon(\omega)} + 1}. \tag{9}$$

In our simulations we used the same parameters for the permittivity as those in Smith & Manogue [2] ($A = 100$, $\omega_0 = 10$, and $\Gamma = 1$). In Figs. 2 (a) and (b) we show the exact real and imaginary parts of the complex reflection coefficient calculated from Eqs. (8, 9), and those inverted with S&M (1) and SSKK (7) dispersion relations for the case $n=1$ without data extrapolation beyond the integration interval, which is shown by the arrows on the horizontal axis. Obviously the results of data inversion are not satisfying. The abrupt change of the relevant curves in Fig. 2 at the initial and final points of the spectrum is due to the logarithmic divergence of the dispersion integrals.

In the case of $n=10$ from almost perfect match between the original and the calculated $\text{Re}\{r(\omega)\}$ and $\text{Im}\{r(\omega)\}$ are obtained, however, one may observe from Figs. 3 (a) and (b) that abrupt change at the borders of the spectral interval persists. It is remarkable that for high $n$, the S&M and SSKK dispersion relations produce practically the same result. Moreover, S&M dispersion relations greatly simplify analysis allowing one to get rid of the anchor point/points. The numerical simulation shows that in the noiseless case, a good correspondence between reflection coefficients calculated from (1) and (7) can be achieved for $n=10$. The utilization of the phase and logarithm of the reflection coefficient, i.e. conventional K-K technique, gives poorer results. In the presence of the noise, lower value of $n$ may be advisable because the contribution of the noise may be overestimated when integration of a high power of reflection coefficient in the spectra analysis. We remark that in the case of SSKK we have two singular points on the integration interval, that have to be reached in a symmetric manner from the left and right- hand side in order to guarantee the success of the numerical integration.

Fortunately in linear optical reflection spectroscopy, one usually can record data at rather wide range so the reliability of the data inversions becomes better. However, in nonlinear optical spectroscopy, the measurement range is usually quite narrow thus dispersion relations analogous to those presented above may be useful in the analysis on nonlinear optical spectra.

It is obvious from Figs. 2 & 3 that we have real and imaginary parts that present symmetric functions and the average value of the real part of the reflection coefficient is zero. Unfortunately, in many cases the situation is not so simple. There may appear offset in the real part so that the average is no more zero. Such an offset has a consequence on the analysis based on S &M and SSKK dispersion relations, as it will be shown in next section.

Before closing this section we would like to remark that Goplen *et al* [13] have derived a band-limited K-K relation for extracting the real refractive index from transmission and attenuated total reflection data. The relevant iterative algorithm has been developed by Nitsche & Fritz [14]. Our method differs from these methods because we want to increase the convergence of the dispersion integrals.

## V.     Testing the consistency of the THz reflection spectra

In THz time-domain reflection spectroscopy, one can obtain simultaneously both the amplitude and the phase of the complex reflection coefficient from the time-domain waveform of the reflected electric field. These measurements conventionally include a comparison of the signal reflected from the sample and reference. Since it is virtually impossible to adjust the sample and the reference exactly at the same position the measured phase carries a frequency dependent misplacement error. While the amplitude of the complex reflection coefficient can be considered to be correct, since it is measured at the far-field region, its frequency dependent real and imaginary parts are not correct. In such a situation extraction of the correct complex refractive index of opaque sample at the terahertz spectral range is not reliable. This constitutes the sample-misplacement problem, which is common in the THz time-domain spectroscopy in reflection geometry. Fortunately, it is possible to correct the phase using maximum entropy model (MEM) [3]. The consistency of the corrected data (i.e. whether real and imaginary parts of the reflection coefficient fulfill the causality principle) should be checked using dispersion theory. However, the conventional K-K technique involves logarithm of reflectance, i.e. the dispersion relations may become invalid for particular choice of the angle of incidence where the complex reflection coefficient may be zero at a given frequency. In such a case we have to generalize K-K relation using so-called Blaschke product [1], which requires the knowledge of the complex zeros of the reflection coefficient. The dispersion relations for the powers of the reflection coefficient, which we are using here, allow us to avoid such pathological difficulty of the standard K- K analysis.

Next we demonstrate the performance of the S&M and SSKK dispersion relations for the reflection spectra of n-type, undoped (100) InAs wafer in the spectral range 0.5–2.5 THz. The experimental setup and detailed description of the sample were reported in Ref. [15]. The data was obtained using p-polarized THz radiation at oblique angle of incidence of 35 degrees. The reference was aluminum and it is assumed that its reflection coefficient is unity in the THz spectral range of interest.

The measured in [15] spectra of the amplitude and the phase of the reflection are shown in Fig. 4. Due to the sample misplacement there is an error in the phase shown in Fig. 4. In THz spectral region, the permittivity can be obtained in terms of the classical Drude model that yields

$$\varepsilon(\omega) = \varepsilon_b \left(1 + \frac{A}{\omega_0^2 - \omega^2 - i\Gamma\omega}\right) \qquad (10)$$

where $\varepsilon_b$ and $\Gamma$ are the background permittivity and damping factor, respectively. In Fig. 4 we show also the amplitude and phase calculated with Drude model and Fresnel's formula for reflection of p-polarized light at oblique incidence. One can observe in Fig. 4 that the measured and the calculated amplitude of the reflection coefficient match pretty well, while the experimental and the calculated phase of the reflection coefficient do not correspond to each other. The reason is the sample misplacement, which can be removed by using MEM technique. The phase corrected by MEM matches well with phase obtained using the Drude model for permittivity [3].

In Fig. 5 we show the erroneous real and imaginary parts of the reflection coefficient calculated using uncorrected data of Fig. 4. As an example, in Fig. 5 we show also the results of the S&M and SSKK analysis when inverting the incorrect data for the case $n = 1$. Now it should be emphasized that we have utilized pairs of dispersion relations, whereas in conventional K-K analysis only the other partner of the K-K relations is exploited. In other words here we invert the measured real part and imaginary parts. If the measured data are not correct then there may appear inconsistency between measured and calculated data. In Fig. 5 we observe that the calculated curves depart to a great extent from the experimental ones. Hence we may draw a tentative conclusion that the measured THz data is not correct.

We have already mentioned that for the validity of the conventional K-K-, S&M- and MSKK relations the true complex reflection coefficient has to be a holomorphic function in the upper half of complex plane. It must also have sufficient decay for high complex frequency and fulfill a symmetry relation due to the parity of the complex reflection coefficient. The phase error due to sample displacement can be expressed as

$$\Delta\varphi = \frac{\Delta L}{c}\omega, \qquad (11)$$

where $\Delta L$ is difference in the optical path between the sample and reference, $c$ is the speed of light in the vacuum. Since the spot size of the incident THz beam on the sample surface is about 0.3 mm in the case of present data, the ratio $\Delta L/c$ is a constant. One can expect that the spectra of the real and imaginary parts of the reflection coefficient are subject to an error when the phase error is large enough as in the case of Figs. 4 and 5. Now the erroneous reflection coefficient can be given as follows:

$$r_{err}(\omega) = |r_{err}(\omega)| \exp\left\{i\left(\varphi(\omega) + \frac{\Delta L}{c}\omega\right)\right\} \quad (12)$$

This function is holomorphic in the upper half plane and fulfills the symmetry relations imposed on the true reflection coefficient. However, if we take a logarithm of this function (12) in order to use the conventional K-K relations we observe that the imaginary part tends to infinity as the frequency tends to infinity. Thus we are no more working with the minimum phase problem [16, 17] and the logarithm is a multi-valued function, which however can be made single valued using the concept of Riemann surface [18]. Next we study function (12) itself for the case $\Delta L/c < 0$. When we extend the exponent function, involving the incorrect phase in Eq. (12), into the upper half plane, $\tilde{\omega} = u + iv$, we observe that we break the reflection coefficient property of tending to zero as the modulus of the complex frequency tends to infinity. Indeed, the function $\exp\{\Delta L v/c\}$ blows up as $v$ tends to infinity. Hence, the validity of K-K, S &M and MSKK is broken, since the asymptotic fall-off of the reflection coefficient in the upper half plane is no more valid. This is because the Jordan's lemma needed in the complex contour integration of derivation of K-K or modified K-K relations is no more valid (see Appendix C in Ref. 4). Hence, one may get strange data inverted curves as those in Fig. 5, which correspond to the case $\Delta L/c < 0$. We may try to avoid this problem of asymptotic behavior of a function. In principle this is possible by deriving dispersion relations for the negative optical path length in the lower half plane. However, then we have to face the problem of poles of the reflection coefficient that are located in the lower half plane. Obviously if $\Delta L/c > 0$ we find a function that has the appropriate asymptotic fall-off at high frequency in the upper half plane. Then one may expect better match with erroneous experimental and inverted curves since the problem of asymptotic fall of the complex contour integration is not present. Indeed, this is the case. In our simulations we have observed that neither S&M nor SSKK relations cannot distillate the error in the case $\Delta L/c > 0$! It is interesting that the validity or invalidity of the K-K-, SSKK- S&M dispersion relations, which are given in this study using the exponential expression of the reflection coefficient, depends on the geometry of the experimental set up. As we already discussed this has purely mathematical origin, and there is no break up of the principle of causality. In THz experiments one can choose the geometry so that the offset of the incident wave corresponds to negative alpha. Thus the methods described in this paper are applicable.

Naturally when we find the phase error and correct the data accordingly, we arrive at agreement between the measured and inverted data. Such a correction of the measured phase is possible using the MEM procedure. However, importantly MEM procedure has no physical basis but it merely employs methods of the information theory. Therefore, it is important to test whether the MEM procedure allows us to obtain correct phase or not. This is the point where the S&M and SSKK dispersion relations also should be employed, in addition to the indication of the negative misplacement upon application of dispersion relations to erroneous real and imaginary parts.

Real part of the reflection coefficient calculated by S&M and SSKK dispersion relations for $n = 1$ using the corrected experimental data is presented in Fig. 6. It is interesting to observe from Fig. 6 that the SSKK relation reproduces the real part from imaginary part pretty well, whereas in the curve obtained by S&M there is an offset. However, it is important to calculate the real part using both S&M and SSKK dispersion relations, as it will be described below. An approximation of this offset can be obtained by averaging the difference of SSKK and S&M curves over the relevant spectral range. It can be shown that if we shift the S&M curve in Fig. 6 by the offset value we get a little bit better approximation for $\text{Re}\{r(\omega)\}$ in this S&M case than in the SSKK case. However, if we just substitute the obtained correct real part of the reflection coefficient into relevant dispersion relation we get absurd values for $\text{Im}\{r(\omega)\}$ both with the S&M and SSKK analysis. The reason is that in order to calculate the imaginary part of the reflection coefficient we have to take into account the finite spectral interval of the experimental data. Specifically, the derivation of the dispersion relations involves an identity

$$P\int_0^\infty \frac{C}{\omega^2 - \omega'^2} d\omega = 0, \tag{17}$$

where $C$ is a constant. In the case of a finite spectral range, the integration is determined by the frequency interval of the experiment and, correspondingly, identity (17) can no longer be employed. It can be shown that in such a case we arrive at finite, frequency-dependent offset, $r_{offset}$, to the real part of reflection coefficient that has to be used in the dispersion relations. Thus in order to calculate the imaginary part of the reflection coefficient one needs to substitute $r - r_{offset}$ into the S&M and SSKK dispersion relations.

In order to obtain $r_{offset}$ we need first to employ the S&M dispersion relation for $n=1$. After the subtraction of the offset from the real part of the reflection coefficient the imaginary part is calculated for high power $n$. Naturally high power technique can be applied also to test the real part. In Fig. 7 we show the curves for $n=10$ for $\text{Re}\{(r-r_{offset})^{10}\}$ and $\text{Im}\{(r-r_{offset})^{10}\}$, which were calculated using appropriate S&M and SSKK relations and the Drude model for fitting the experimental data. The curves match very well with the exact ones. Obviously high power of reflection coefficient is effective in testing band-limited data. One can freely select which to use S&M or SSKK (or both) for high power $n$. The real and imaginary parts of the complex reflection coefficient can be calculated by taking $10^{th}$ root of the data of Fig. 7.

## VI. Conclusions

In conclusion, we have demonstrated that S&M and SSKK dispersion relations for the power of complex reflection coefficient provide us with powerful tools in order to improve data inversion at a relatively narrow spectral range. We compared the S&M and SSKK analyses, and observed that in the present cases S&M performs as good as SSKK that needs information at the anchor point. As an application the developed analysis can be applied to verify the consistency of the experimental data in terahertz spectroscopy, for the case of negative misplacement, when both amplitude and phase of the reflection coefficient can be measured, and the success of phase correction by MEM procedure. The MEM procedure is a valuable tool in phase correction, which is needed in solving the problem of sample misplacement.

We remark that in principle it is also possible to use MEM & dispersion analysis as a feedback system to monitor and find the correct position of the sample in the THz - experimental setup. We believe that this kind of approach is valid for the problems of another kind of experimental ambiguities. Finally the developed technique can be exploited for testing the consistency of the experimental data obtained in ellipsometry and nonlinear optical spectroscopy measurements.

# Figure captions

Fig. 1. Contour of complex integration.

Fig. 2. Real (a) and imaginary (b) part of the Lorentzian complex reflection coefficient calculated using the S&M (dashed line) and SSKK (dotted line) relations at $n$ = 1. Solid lines represent $\text{Re}\{r(\omega)\}$ and $\text{Im}\{r(\omega)\}$ given by Eqs. (8, 9).

Fig. 3. Real (a) and imaginary (b) part of the Lorentzian complex reflection coefficient calculated using the S&M (dashed line) and SSKK (dotted line) dispersion relations at $n$ = 10. Solid lines present exact curves.

Fig. 4. Upper panel: measured amplitude of InAs at terahertz range (filled circles) and amplitude obtained from Drude model (solid line). Lower panel: measured phase (filled circles), phase obtained from Drude model (solid line) and corrected phase of the complex reflection coefficient obtained by MEM (filled triangles).

Fig. 5. Incorrect real and imaginary parts of the reflection coefficient (solid lines). The curves obtained by the S&M (dashed line) and SSKK (dotted line) analysis were calculated using the experimental data (solid lines) for the case $n$ = 1.

Fig. 6. Real part of the reflection coefficient of InAs calculated from S&M (dashed line) and SSKK (dotted line) dispersion relations for the case $n = 1$. The solid line presents the real part, which was obtained after phase correction by MEM.

Fig. 7. Re$\{(r-r_{offset})^{10}\}$ (a) and Im$\{(r-r_{offset})^{10}\}$ (b) of InAs calculated from the S&M (dotted line) and SSKK (dashed line) dispersion relations in the case $n = 10$. The solid lines present the curves obtained using Drude model for InAs. The filled circles correspond to experimental data points.

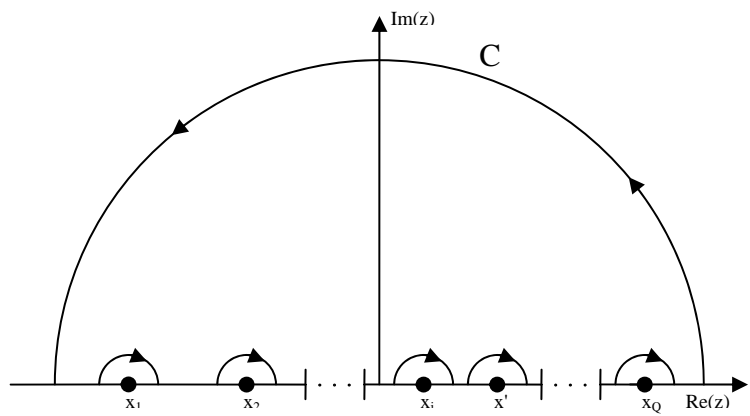

Fig. 1. K.-E. Peiponen *et al*

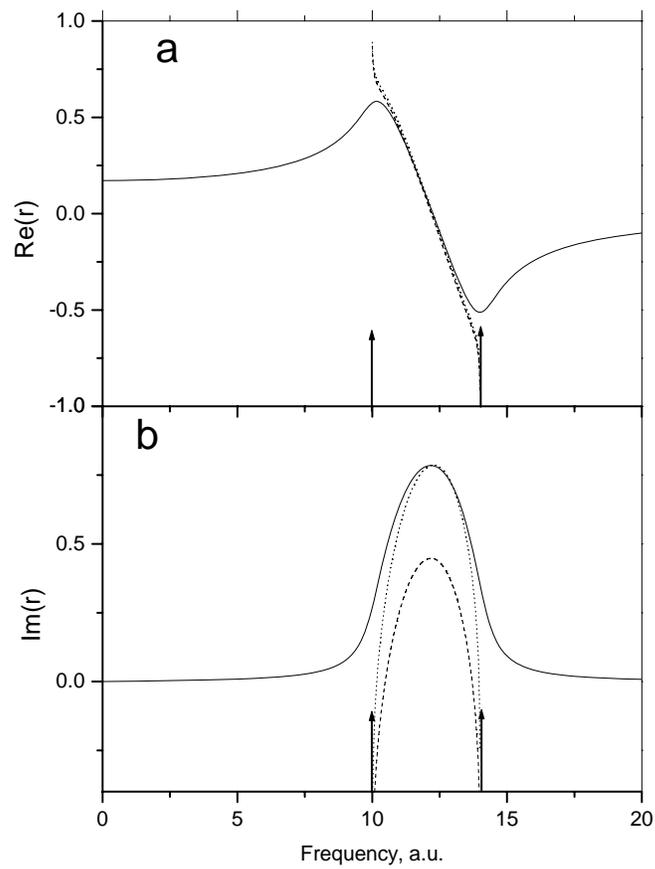

Fig. 2. K.-E. Peiponen *et al*

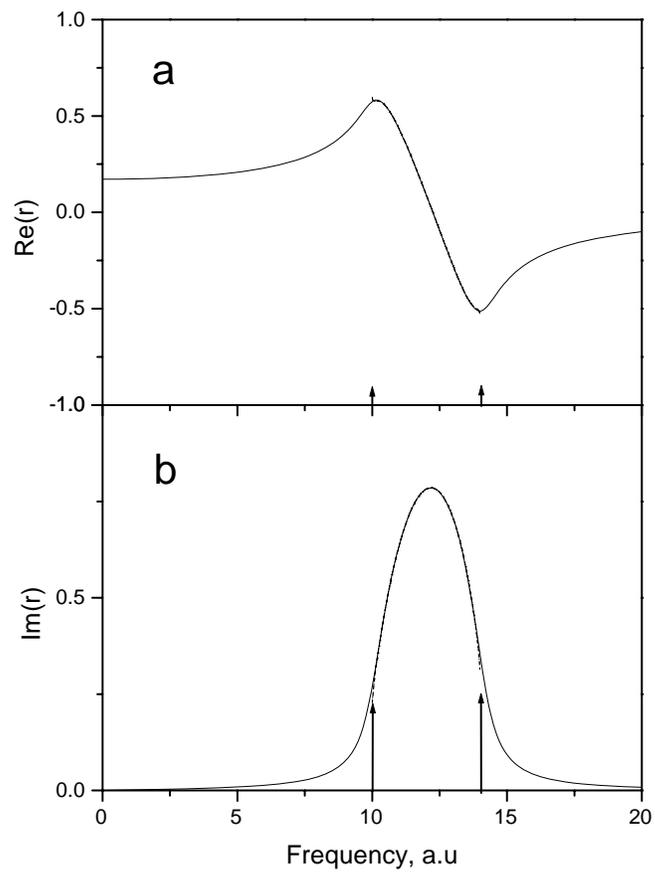

Fig. 3. K.-E. Peiponen *et al*

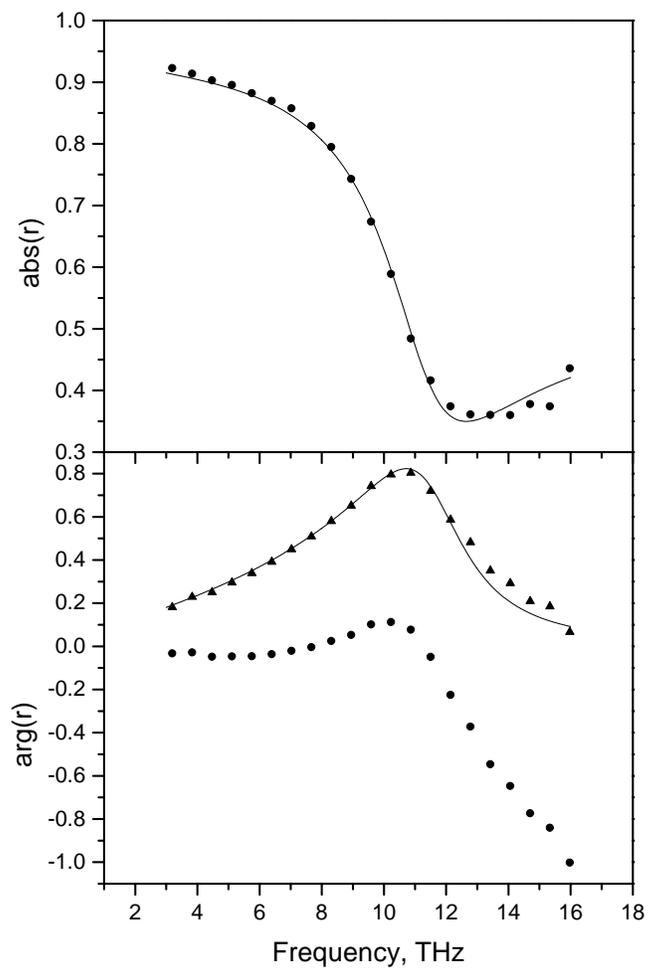

Fig. 4. K.-E. Peiponen *et al*

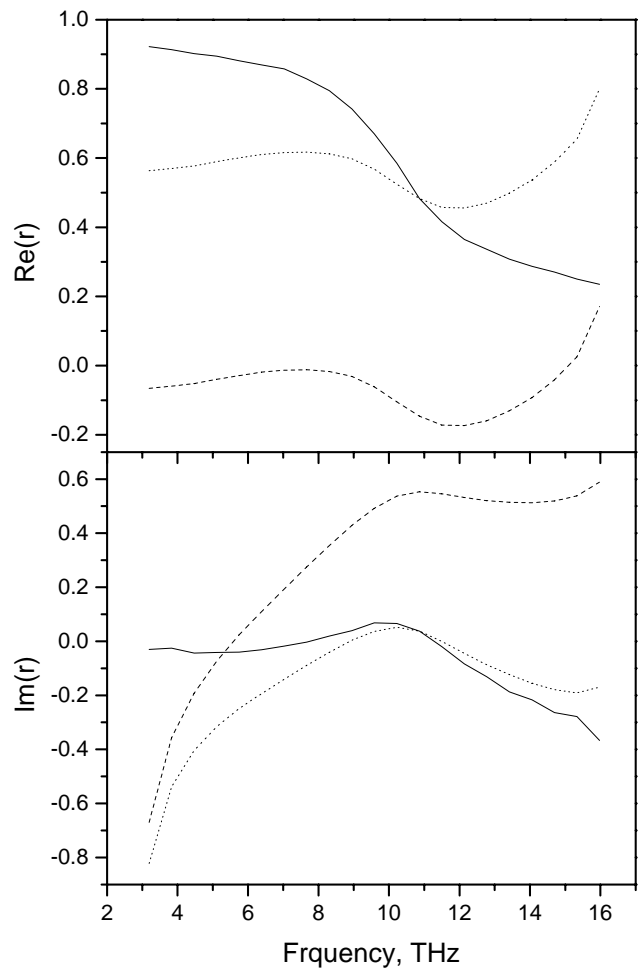

Fig. 5. K.-E. Peiponen *et al*

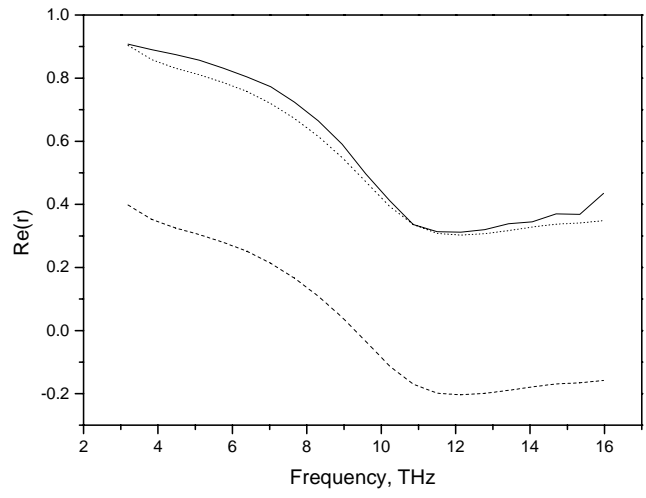

Fig. 6. K.-E. Peiponen *et al*

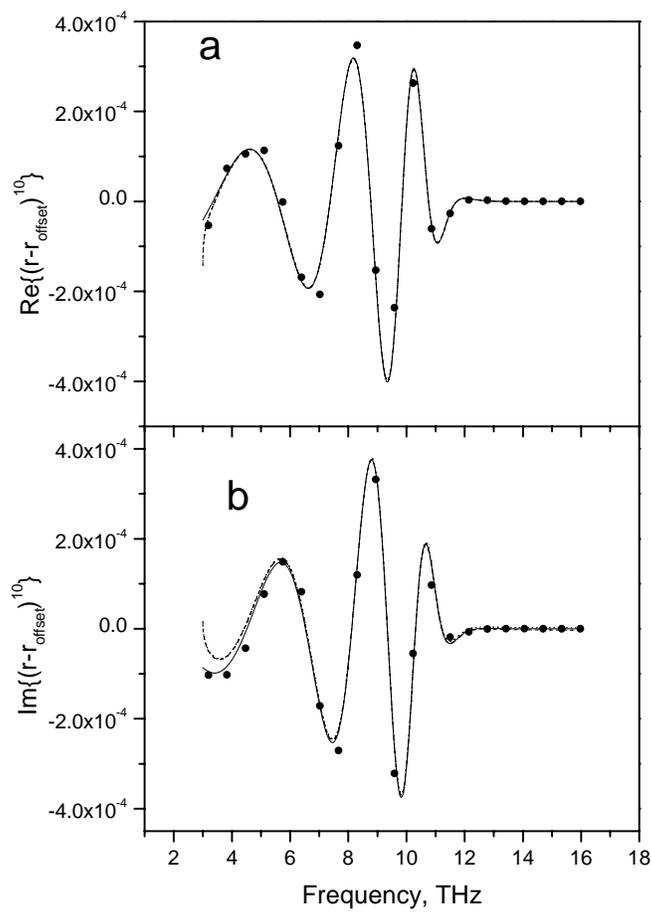

Fig. 7. K.-E. Peiponen *et al*